\documentclass[prl,twocolumn,showpacs,floatfix]{revtex4}
\usepackage{graphics,epsfig,amsfonts,amssymb,amsmath}
\begin{document}

\title{Conductivity anisotropy in the antiferromagnetic state of
iron pnictides}
\author{B. Valenzuela}
\email{belenv@icmm.csic.es}
\affiliation{Instituto de Ciencia de Materiales de Madrid,
ICMM-CSIC, Cantoblanco, E-28049 Madrid (Spain).}
\author{E. Bascones}
\email{leni@icmm.csic.es}
\affiliation{Instituto de Ciencia de Materiales de Madrid,
ICMM-CSIC, Cantoblanco, E-28049 Madrid (Spain).}
\author{M.J. Calder\'on}
\email{calderon@icmm.csic.es}
\affiliation{Instituto de Ciencia de Materiales de Madrid,
ICMM-CSIC, Cantoblanco, E-28049 Madrid (Spain).}
\date{\today}
\begin{abstract}
Recent experiments on iron pnictides have uncovered a 
large in-plane resistivity anisotropy with a surprising result: the system 
conducts better in the antiferromagnetic $x$ direction 
than in the ferromagnetic $y$ direction. 
We address this problem by calculating the ratio of the Drude weight along the 
$x$ and $y$ directions, $D_x/D_y$, for the mean-field $\bf{Q}=(\pi,0)$
magnetic phase diagram of a five-band model for the undoped pnictides.   
We find that $D_x/D_y$ ranges between $0.2 < D_x/D_y < 1.7$
for different interaction parameters. Large values of the orbital ordering favor 
an anisotropy opposite to the one found experimentally. 
On the other hand, $D_x/D_y$ is strongly dependent on the topology and morphology 
of the reconstructed Fermi surface. Our results point against
orbital ordering as the origin of the observed conductivity anisotropy, which
may be ascribed to the anisotropy of the Fermi velocity.
\end{abstract}
\maketitle

Understanding the 
magnetic state of iron pnictides is
probably the starting point to understand the origin of
superconductivity in these compounds.  The magnetic state is metallic, 
in contrast to the Mott-insulating behavior of cuprates, and it presents 
$Q=(\pi,0)$ columnar ordering: antiferromagnetic in the
$x$ direction and ferromagnetic in the $y$ direction. The
measured magnetic moment is unexpectedly small and
a structural transition occurs at or very close to the magnetic transition~\cite{zhao08}.

Recent experiments~\cite{chu10-1,chu10-2,mazin10} have measured an
anisotropic resistivity in the magnetic state with unexpected results: 
The resistivity is smaller in the antiferromagnetic $x$ direction than in the
ferromagnetic $y$ direction. A priori the opposite would be expected.
The anisotropy extends to finite frequencies as seen in optical
conductivity experiments~\cite{degiorgi10,postersns}. 
Surprisingly, the resistivity anisotropy grows with doping and is largest 
close to the border of the antiferromagnetic phase when the magnetism  and the 
structural distortion are weaker. The anisotropy cannot be explained by 
the small changes in the lattice constants produced by the structural 
transition as the lattice constant is larger in the direction with larger 
conductivity~\cite{huang08}. It can neither 
 be  ascribed to the scattering rate of the carriers which has been observed
to increase  along 
the $x$ direction and decrease along the $y$ direction in the magnetic state~\cite{degiorgi10}, as expected on the basis of scattering by spin
fluctuations~\cite{chen_deveraux10}.

The anisotropy of the magnetic state has also been reported in other experiments:(i) Neutron scattering measurements~\cite{zhaonatphys09}
were described with anisotropic 
exchange constants; 
(ii) angle resolved photoemission spectroscopy (ARPES) finds
a predominant $zx$ orbital component at the Fermi surface at 
$\Gamma$~\cite{shimojima10}; and
(iii) spectroscopic imaging scanning tunneling microscopy
experiments~\cite{sciencedavis10} were interpreted in terms of
  one-dimensional bands and of electronic nanostructures  
aligned along the $x$-axis.

The origin of the 
anisotropy is at the center of a hot debate. 
Orbital 
ordering (OO)
in the $d_{zx}$ and $d_{yz}$ orbitals
has been proposed~\cite{Castro-Neto,Kruger09,sciencedavis10,singh-09,degiorgi10,chen_deveraux10,yin_kotliar10,leeyinku09,lv_phillips10,yinleeku10} and found in the magnetic state in several 
theoretical approaches~\cite{Kruger09,daghofer10,leeyinku09,nosotrasprl10}. 
On the other hand, the $(\pi,0)$-antiferromagnetism breaks the tetragonal
symmetry and could induce anisotropy without invoking other electronic
sources \cite{mazin10,mazin_comment,daghofer10-2}. 
Within this context it is crucial to 
uncover the origin of the experimentally observed anisotropies.

\begin{figure*}
\leavevmode
\includegraphics[clip,width=\textwidth]{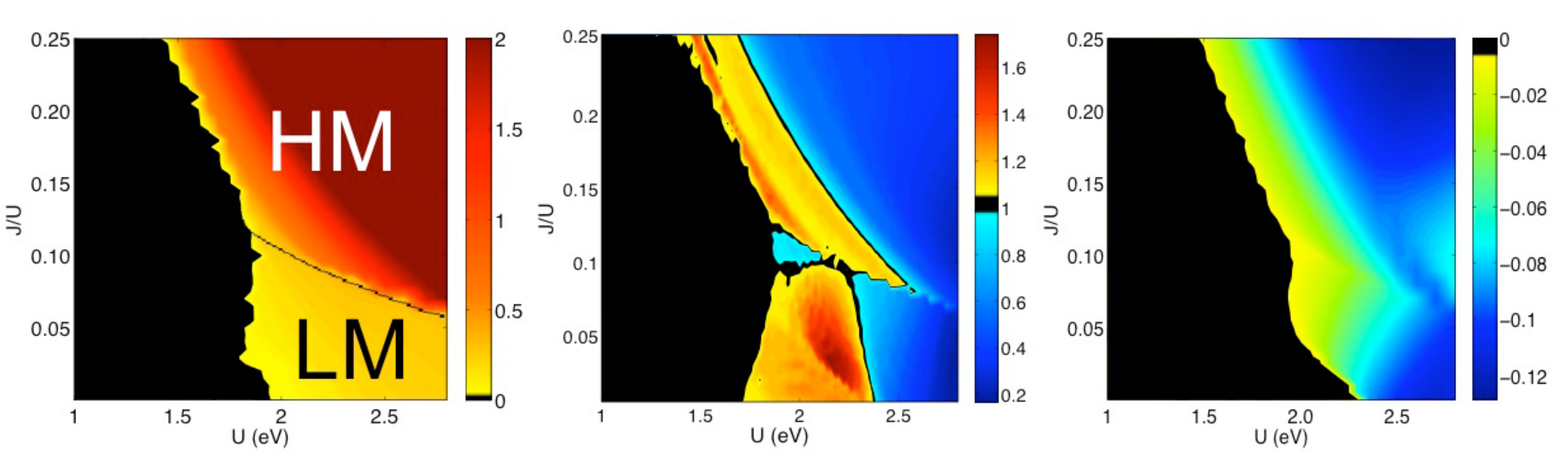}
\caption{ (Color online) (a) $U$ vs $J/U$ magnetic phase diagram superposed to 
the magnetization showing the paramagnetic (black), high moment (HM) and low 
moment (LM) magnetic phases.
 The magnetic moment reaches values as high as 3.4 $\mu_B$. The color scale 
emphasizes the region with magnetic moments  smaller than 2 $\mu_B$.  
(b) Drude ratio with largest values of $D_x/D_y>1$ corresponding to small 
values of the magnetization. 
(c) Orbital ordering $n_{yz} - n_{zx}$. It correlates well with the difference
between the partial density of states of $d_{yz}$ and $d_{zx}$ at the Fermi
level, see supplementary material.} 
\label{fig:maps}
\end{figure*}

In a former work we have calculated the mean field phase diagram of
the magnetic state~\cite{nosotrasprl10} for a five orbital
model~\cite{nosotrasprb09}. 
Two different metallic states have been found: the so-called  
high moment (HM) and low moment (LM) states.
In the HM state all the orbital magnetizations point 
in the same direction leading to a large local magnetic moment,
except for the smallest values of $U$ which give a magnetic solution [see
Fig.~\ref{fig:maps}(a)].  
On the contrary, the LM state consists of antiparallel orbital 
magnetizations which add
up to a small total magnetization.  This state, stabilized 
by  the anisotropy of the interorbital exchange interactions, 
has also been found in   other mean-field~\cite{Dagotto10arpes} and  LDA+U~\cite{cricchio09} calculations.

OO defined in terms of a different occupation of $d_{yz}$ and
$d_{zx}$, $n_{yz}-n_{zx}$, is present in
both the LM and HM states [see Fig.~\ref{fig:maps}(c)].       
The orbital polarization of the Fermi surface shows larger $zx$ orbital
component around $\Gamma$~\cite{nosotrasprl10} in agreement 
with experiments~\cite{shimojima10}. Whether the $zx$ polarization of the Fermi
surface requires strong OO, i.e. finite $n_{yz}-n_{zx}$, or it is just a consequence
of the orbital reorganization at the Fermi surface due to the coupling between
spin and orbitals is currently under discussion~\cite{chen_deveraux10,daghofer10-2}.  
Furthermore, we estimated the anisotropy of the exchange interactions 
by mapping the multi-orbital mean field state to a Heisenberg model, and 
found a large anisotropy only in the LM state showing no correlation with 
the magnitude of the OO~\cite{nosotrasprl10}.

In this letter we calculate the ratio of the Drude weight
$D_x/D_y$ and analyze how it correlates with the presence of OO
for the metallic region of a five-orbital $Q=(\pi,0)$ 
mean-field phase diagram to unveil the possible source of the transport 
anisotropy observed in pnictides~\cite{chu10-1,chu10-2,mazin10,degiorgi10,postersns}.
We find a wide range of values for this ratio ($0.2 < D_x/D_y < 1.7$) 
depending on the value of the interaction parameters.
Larger OO appears together with larger conductivity in the ferromagnetic 
direction, therefore it cannot be responsible for the observed anisotropy.
 We show that the magnitude of the Drude anisotropy is related  
 to the topology and morphology of the Fermi surface. 
The  observed conductivity anisotropy 
 may be ascribed to the Fermi velocity, as in Ref.~\cite{mazin10}.

The 5-orbital interacting model Hamiltonian  and mean-field $U$ vs $J/U$
  $\bf{Q}=(\pi,0)$ magnetic phase diagram were
described in detail in Ref.~\cite{nosotrasprl10}. 
As detailed in Ref.~[\onlinecite{nosotrasprb09}], 
both direct Fe-Fe and indirect (via As) hoppings, calculated within the Slater-Koster framework~\cite{slater54}, determine the magnitude of the hopping amplitudes
which depend on the angle $\alpha$ formed by the Fe-As bonds and the
Fe-plane. Here we use $\alpha=35.3^o$ corresponding to a regular As
tetrahedra. Energy units are $\sim 1 eV$. The interactions include the intraorbital Hubbard $U$, the Hund's
coupling $J$, and the interobital $U'=U-2J$. 
The Hamiltonian is solved at the mean-field level keeping only the spin and 
orbital-diagonal average terms.  
In Fig.~\ref{fig:maps}(a) we reproduce the phase diagram in the metallic 
region ($U \leq 2.8$ eV) superposed to the resulting magnetization.  
As discussed above, HM and LM solutions arise for high and 
low values of $J$ and both show OO,  see Fig.~\ref{fig:maps}(c). OO observed
  here has the same sign as the one obtained in LDA+U~\cite{leeyinku09} and
  LDA+DMFT~\cite{yin_kotliar10}.

The Drude weight involves the carrier velocity,  the
  spectral weight, and the scattering rate. Experimentally, the
  scattering rate is anisotropic, but with an anisotropy opposite to that of
  the resistivity~\cite{degiorgi10}. Here 
we neglect the influence of the scattering rate. 
This approximation allows us to determine the role of the Fermi velocity in 
the observed transport anisotropy. We calculate 
\begin{equation}
 D_x/D_y=\frac{\sum_{{\bf k},n}v^2_{x}({\bf k},n)
 g({\bf k},n)\delta (\epsilon_n({\bf k})-E_F))}{\sum_{{\bf k},n}v^2_{y}({\bf
   k},n) g({\bf k},n)\delta(\epsilon_n({\bf k})-E_F)}
\end{equation}
with $v_{i}$ the
velocity in the direction $i=x,y$, $n$ the band index, $g$ the 
spectral weight, $\epsilon_n$ the energy of the band $n$, $\delta$ the Dirac function, $E_F$ the Fermi energy, and the sum in $\bf k$  restricted to the 
Fermi surface. The result for the Drude weight ratio is 
shown in Fig.~\ref{fig:maps}(b). 
In the paramagnetic (PM) region, 
black region in the magnetic phase diagram Fig.~\ref{fig:maps}(a), $D_x/D_y=1$ 
as expected (discrepancies along the boundary region might 
be due to numerical precision). The magnetic phases show anisotropy in the
Drude weight with ratios $0.2< D_x/D_y <1.7$. $D_x/D_y> 1$, consistent with
the experimentally observed transport
anisotropy~\cite{chu10-1,chu10-2,mazin10}, is observed within a great part of
the  LM region for intermediate values of the on-site interaction well suited
for describing pnictides. In the  HM phase, $D_x/D_y>1$
is restricted to a region close to the PM phase, corresponding to
the smallest
values of the magnetic moment. 
Interestingly,
in a wide part of the region with $D_x/D_y>1$ an increase of 
the anisotropy correlates
with a decrease of the magnetic moment. On the other hand $D_x/D_y>1$ happens 
to be when the OO is 
small ($|n_{yz} - n_{zx}|\lesssim 0.05$). 
Note that although there is not a perfect one to one 
correspondence between the anisotropy of the Drude weight 
[Fig.~\ref{fig:maps}(b)] and the OO [Fig.~\ref{fig:maps}(c)],
$D_x/D_y$ decreases with increasing OO in most part of the phase diagram. That
is, OO seems to favor a transport anisotropy opposite to the one reported
experimentally~\cite{mazin10}. This finding rules out OO as the origin of the 
observed resistivity anisotropy.

\begin{figure}
\leavevmode
\includegraphics[clip,width=0.45\textwidth]{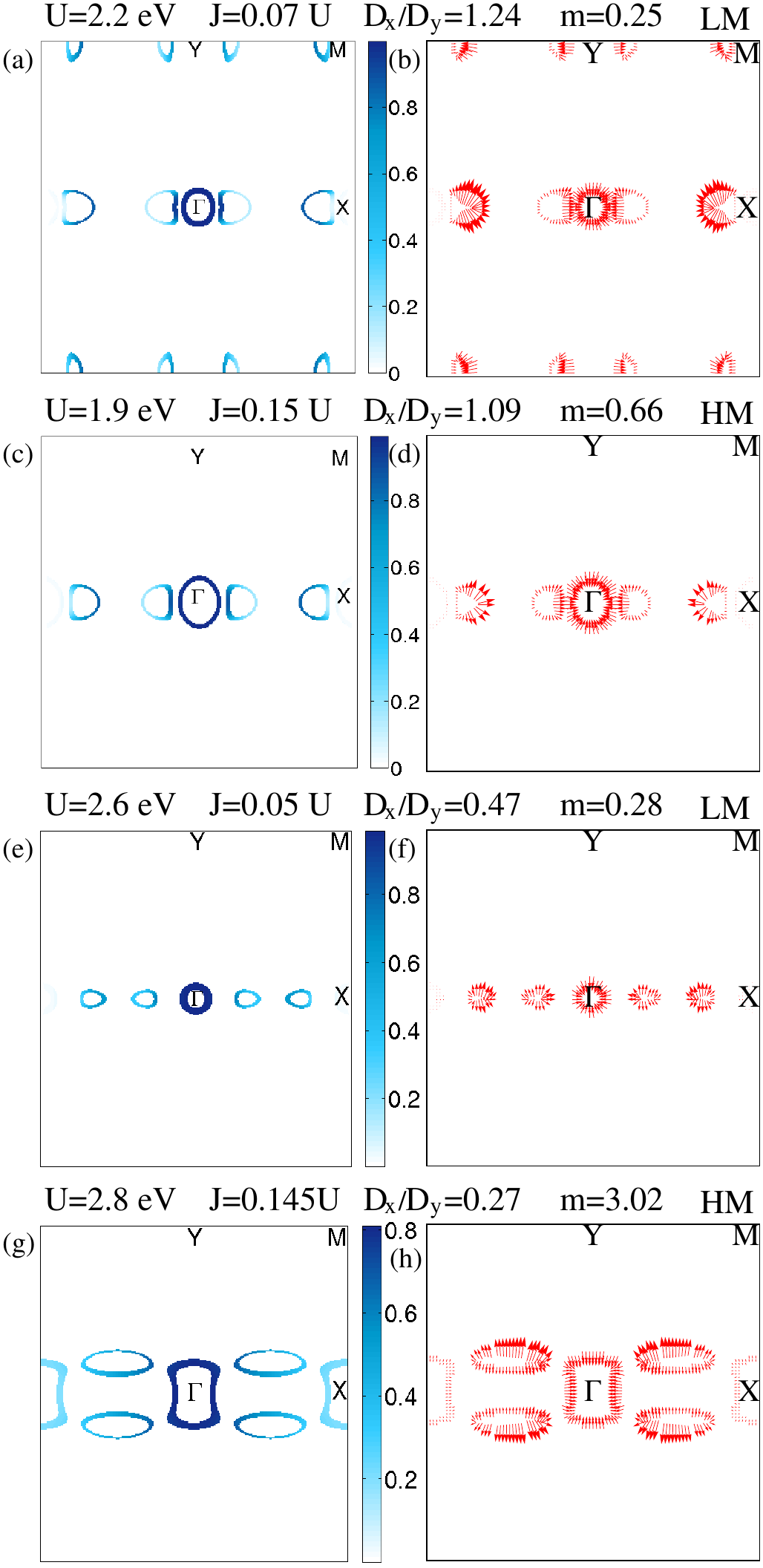}
\caption{ (Color online) Fermi surfaces (left) and velocities (right) at
  different 
points of the phase diagram  in the extended (Fe) Brillouin zone. The
  Fermi surfaces are weighted by their spectral weight. The
  corresponding $U$ and $J$ values, as well as the obtained 
  anisotropy of the Drude ratio, magnetic moment in Bohr magnetons 
and the magnetic state appear on
  top of each figure.  
}
\label{fig:fsvel}
\end{figure}

\begin{figure*}
\leavevmode
\includegraphics[clip,width=0.85\textwidth]{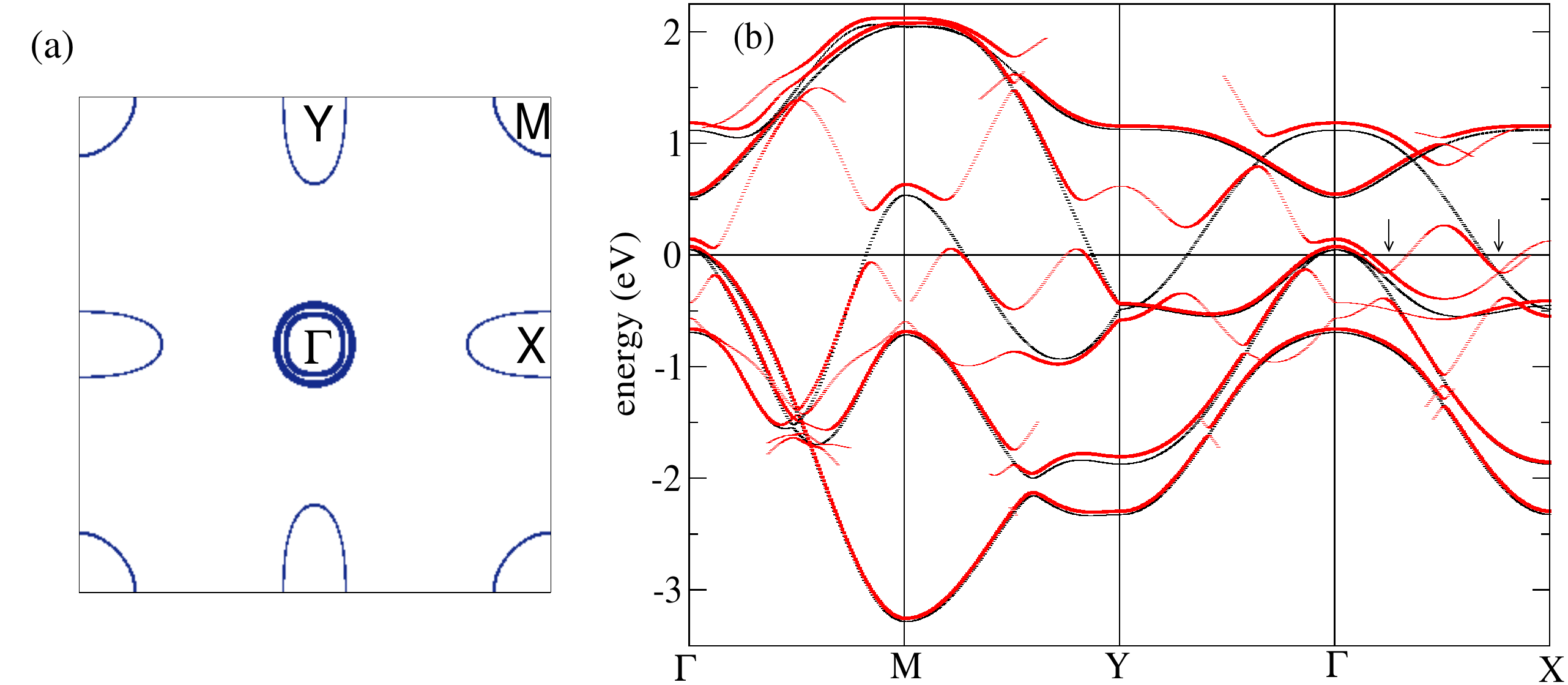}
\caption{ (Color online) (a) Fermi surface in the paramagnetic state
  for $U,J=0$. (b) Band structure for $U=2$ eV and $J/U=0.07$ corresponding 
  to the LM state, in red. The linewidth gives a measure of the
  spectral weight. The paramagnetic bands, in black, are
  shown as a reference. The position of the Dirac cones along $\Gamma X$ is
  marked with arrows.} 
\label{fig:bandasyfs}
\end{figure*}

We now analyze the Drude weight results on the light of the 
 shape of the  reconstructed Fermi surface and the velocity vector on 
it~\cite{mazin10}. 
These are
represented in Fig.~\ref{fig:fsvel} in the extended (Fe) Brillouin zone
for different points of the phase diagram corresponding to different values of
the ratio $D_x/D_y$. We have chosen Fermi surfaces representative of a wide
area of the phase diagram, while other topologies have also been found. 
We observe a strong dependence of the morphology and topology of the Fermi
surfaces on the interaction parameters. These variations, which have a
correspondence with the calculated Drude weight ratio, can be understood
starting from the PM Fermi surface  in Fig.~\ref{fig:bandasyfs}(a) 
and the plot of the bands in Fig.~\ref{fig:bandasyfs}(b). 
The bands are shown for the PM state (in black) and the LM state (in red) 
for $U=2.2$ eV and $J/U=0.07$. 

The PM Fermi surface has hole pockets in $\Gamma$ and $M$, and
electron pockets in $X$ and $Y$. In the magnetic states, a gap opens at $M$
and $Y$ near $E_F$ but, for small values of the magnetic moment, 
 there are still pockets in $M \to Y$ 
though considerably reduced in size with respect to the
PM phase, see Fig.~\ref{fig:bandasyfs}. This is illustrated in
Fig.~\ref{fig:fsvel}(a) for  $U=2.2$ eV and $J/U=0.07$ (corresponding to the 
 LM state) and is also representative of the 
HM state with a small value of the magnetization (region close to the
PM phase). 
This topology of the Fermi surface corresponds to the largest values 
of the anisotropy $D_x/D_y > 1.2$. A visual inspection of the velocity vectors
at $k_F$ in Fig.~\ref{fig:fsvel}(b) shows that the $v_x$ component of the velocity dominates in this case,
with a large contribution from the pockets at $M \to Y$. $D_x/D_y >1$ 
appears also at points of the phase diagram with the Fermi surface represented 
in Fig.~\ref{fig:fsvel}
(c) for $U=1.9$ eV and $J/U=0.15$. The main difference between
Figs.~\ref{fig:fsvel}(a) and (c) 
is the disappearance of the pockets at $M \to Y$ in the latter case while the
shape of the pockets at $\Gamma$ and along $\Gamma \to X$ is maintained. 
The electron pockets along $\Gamma \to X$ are associated to two 
Dirac cones just below $E_F$ as shown in Fig.~\ref{fig:bandasyfs}. 
The morphology of the
Fermi pockets at 
$\Gamma \to X$ in Figs.~\ref{fig:fsvel}(a) and (c) also supports $D_x/D_y >1$
as evidenced in 
Fig.~\ref{fig:fsvel}(d). It is not straightforward to make a connection
between the anisotropy of the Drude weight and the orbital content of the
Fermi pockets. The pockets close to $Y$ show $zx$ and $xy$
polarization~\cite{nosotrasprb09}, while in 
the Dirac cone electron pocket the $v_x$ contribution comes mostly from the
$d_{yz}$ polarized segment closer to $\Gamma$~\cite{nosotrasprl10}. In fact,
the anisotropy of the Drude conductivity cannot be ascribed to the orbital
reorganization at $E_F$, whose dependence on $U$ and $J$ follows 
that of the OO, see supplementary material. 

Now we turn to Fig.~\ref{fig:fsvel}(e) (for $U=2.6$ eV and $J/U=0.05$). The
topology of the Fermi surface is maintained with respect to
Fig.~\ref{fig:fsvel}(c) but the 
morphology is different: the pockets are more isotropic and those close to $X$
have a slightly stronger $v_y$ component rendering $D_x/D_y \lesssim 1$ [see
Fig.~\ref{fig:fsvel}(f)]. Finally, Fig.~\ref{fig:fsvel}(g) (for $U=2.8$ eV and
$J/U=0.145$) corresponds to a strong anisotropy $D_x/D_y \ll 1$, opposite to
the one observed experimentally. 
The pockets along $\Gamma \to X$ have disappeared. Two ellipsoidal pockets
with a strong $v_y$ component close to this
direction are observed instead, see Fig.~\ref{fig:fsvel}(f).

The Fermi surfaces in  Fig.~\ref{fig:fsvel}(a) and (c), once folded to 
the FeAs Brillouin zone and taken into account the existence of magnetic
domains, resemble the flower-like Fermi surfaces observed in ARPES in the 
magnetic state~\cite{kondo09,he10}.  
From the available experimental results we cannot conclude on the presence or 
absence of Fermi pockets close to $Y$.

In conclusion, we have found that the anisotropy of the Fermi velocity 
which is strongly
related to the morphology and topology of the  reconstructed Fermi surface 
 results in large variations in the ratio of the Drude weight. Depending on
  the shape of the Fermi surface the resistivity can be larger in the
  antiferromagnetic $x$ or the ferromagnetic $y$ direction. 
 Larger orbital ordering correlates with a resistivity anisotropy opposite 
 to the one
  observed, i.e. larger conductivity in the ferromagnetic $y$ direction.
  A larger conductivity in the antiferromagnetic $x$ direction, consistent
  with the experimentally observed conductance anisotropy, corresponds to 
  Fermi surfaces whose shape resembles those observed experimentally.
Values of the anisotropy $D_x/D_y > 1.2$ are achieved when the Fermi surface has pockets at $Y$  in the extended Fe
  Brillouin zone. Pockets around $Y$
are also predicted in ab-initio~\cite{mazin_comment} 
and mean field calculations~\cite{yu09,japonesesmf09}.
The electron pockets originating from the Dirac cones can also govern the
anisotropy when $D_x/D_y> 1$.  The region of the phase diagram with
  $D_x/D_y>1$ shows a small magnetic moment, as observed experimentally.

  We acknowledge funding from Ministerio de Ciencia e Innovaci\'on through 
  Grants No. FIS 2008-00124/FIS, and No. FIS2009-08744 and Ram\'on y Cajal 
  contract, and from CSIC through Grants PIE-200960I033 and PIE-200960I180.

\bibliography{pnictides}

\end{document}